\documentclass[]{article}

\usepackage{graphicx}
\usepackage{afterpage}

\renewcommand{\theequation}{\arabic{section}.\arabic{equation}}

\title{Dimer and trimer fluctuations in the $s=\frac{1}{2}$ transverse $XX$ chain}

\author{Oleg Derzhko$^{1,2}$, 
        Taras Krokhmalskii$^1$, 
        Joachim Stolze$^3$ 
        and 
        Gerhard M\"{u}ller$^4$\\
\small $^1$Institute for Condensed Matter Physics 
           of the National Academy of Sciences of Ukraine,\\ 
\small     1 Svientsitskii Street, L'viv-11, 79011, Ukraine\\
\small $^2$Max-Planck-Institut f\"{u}r Physik komplexer Systeme,\\
\small     N\"{o}thnitzer Stra\ss e 38, 01187 Dresden, Germany\\
\small $^3$Institut f\"{u}r Physik, Universit\"{a}t Dortmund,\\
\small     44221 Dortmund, Germany\\
\small $^4$Department of Physics, University of Rhode Island,\\
\small     Kingston, Rhode Island 02881-0817}

\date{\today}

\begin{document}

\renewcommand\baselinestretch {1.5}
\large\normalsize

\maketitle

\begin{abstract}
  Exact results for the dynamic dimer and trimer structure factors of the
  one-dimensional $s=\frac{1}{2}$ $XX$ model in a transverse magnetic field 
  ($\parallel z$) 
  are presented and discussed in relation to known exact results for the
  dynamic spin structure factors.  In the framework of the Jordan-Wigner
  representation, the accessible spectrum of the dimer fluctuation operator is
  limited to two-fermion excitations whereas that of the trimer fluctuation
  operator involves two-fermion and four-fermion excitations.  The spectral
  boundaries, soft modes, and singularity structure of the four-fermion
  excitation continuum as probed by the dynamic trimer structure factor are
  examined and compared with corresponding properties of the two-fermion
  excitation continuum, as probed by the dynamic dimer and transverse spin
  structure factors.
\end{abstract}

\vspace{3mm}

\noindent
{\bf PACS number(s):}
75.10.-b

\vspace{3mm}


\vspace{5mm}

\noindent
{\bf Postal address:}\\
Dr. Oleg Derzhko (corresponding author)\\
Institute for Condensed Matter Physics
of the National Academy of Sciences of Ukraine,\\
1 Svientsitskii Street, L'viv-11, 79011, Ukraine\\
Tel: (0322) 76 19 78\\
Fax: (0322) 76 11 58\\
E-mail: derzhko@icmp.lviv.ua

\pagebreak

\renewcommand\baselinestretch {1.4}
\large\normalsize

\section{Introduction}
\label{sec1}

\setcounter{equation}{0}

The theoretical and computational study
of frequency-resolved quantum fluctuations and thermal fluctuations
in many-body model systems
is an important area of research for several reasons.
Such fluctuations are observable directly or indirectly
by a host of measuring techniques
used in condensed matter physics and materials science.
The shape of the spectrum
(dispersions, bandwidths, gaps, soft modes, etc.),
the spectral-weight distributions,
and the singularity structure of the dynamic structure factors
as measured or calculated for specific fluctuation operators
yield detailed insights into the state of the material
and reveal important clues about the susceptibility of the system
to phase transitions with order parameters
modeled after the fluctuation operators at hand \cite{01}.

An increasing number of exactly solvable quantum many-body model systems
turn out to be relevant for situations
where certain degrees of freedom of a material
are kinematically constrained to one spatial dimension.
Many such situations offer the most detailed comparisons in many-body dynamics
of exact theoretical results with direct experimental observation \cite{02,03,04}.

Various properties of dynamic spin structure factors
of quantum spin chain models
are observable in quasi-one-dimensional magnetic insulators, for example,
via magnetic neutron scattering \cite{05}.
The properties of dynamic dimer and trimer structure factors,
on the other hand,
are important indicators of structural phase transitions
driven by magnetic interactions such as in spin-Peierls compounds.
Dimer fluctuations are key participants
in phonon-assisted optical absorption processes of magnetic chain compounds
and are thus observable in optical conductivity measurements \cite{06,07,08,09,10}.

Here we consider the exactly solvable $s=\frac{1}{2}$ $XX$ chain with a magnetic field
in the direction transverse to the spin coupling (in spin space)
\cite{11,12}.
The Hamiltonian reads
\begin{eqnarray}
\label{1.01}
H=\sum_{n=1}^N\Omega s_n^z
+\sum_{n=1}^NJ\left(s_n^xs_{n+1}^x+s_n^ys_{n+1}^y\right).
\end{eqnarray}
We set $g\mu_B=1, \hbar=1$, use the exchange constant as the energy unit, and measure
the magnetic field $\Omega$ in units of $J$. 
The Jordan-Wigner transformation to spinless lattice fermions
maintains the bilinear operator structure,
\begin{eqnarray}
\label{1.02}
H=\sum_{n=1}^N\Omega\left(c_n^\dagger c_n-\frac{1}{2}\right)
+\frac{1}{2}\sum_{n=1}^N
J\left(c^\dagger_nc_{n+1}-c_nc^\dagger_{n+1}\right).
\end{eqnarray}
A Fourier transform,
$c_\kappa=N^{-1/2}\sum_{n=1}^N
\exp\left({\mbox{i}}\kappa n\right)c_n$,
brings (\ref{1.02}) into diagonal form:
\begin{eqnarray}
\label{1.03}
H=\sum_{\kappa}
\Lambda_\kappa\left(c_\kappa^\dagger c_\kappa-\frac{1}{2}\right),
\;\;\;
\Lambda_\kappa=\Omega+J\cos\kappa.
\end{eqnarray}
For periodic boundary conditions in (\ref{1.01})
the allowed values of the fermion momenta $\kappa_i$
depend on whether the number $N_f$ of fermions in the system
is even or odd:
$\kappa_i \in \left\{\left(2\pi/N\right)\left(n+\frac{1}{2}\right)\right\}$
if $N_f$ is even
or
$\kappa_i \in \left\{\left(2\pi/N\right)n\right\}$
if $N_f$ is odd.
Fermion momenta within the first Brillouin zone are specified by integers
$n=-N/2,-N/2+1,\ldots,N/2-1$ 
(if $N$ is even)
or
$n=-(N-1)/2,-(N-1)/2+1,\ldots,(N-1)/2$ 
(if $N$ is odd).
The Fermi level in the band $\Lambda_\kappa$
is controlled by the magnetic field $\Omega$.
The number of fermions can vary
between an empty band $(N_f=0)$
and a full band $(N_f=N)$
and is related
to the quantum number $S^z$
(the $z$-component of the total spin)
of the same model in the spin representation:
$S^z=N_f-N/2$.

The dimer and trimer fluctuation operators for the $XX$ model (\ref{1.01}) will
be introduced in Sec.~\ref{sec2}. The two-fermion and four-fermion dynamic
structure factors associated with these fluctuation operators will be discussed
in Secs.~\ref{sec3} and \ref{sec4}, respectively.
Finally,
in Sec.~\ref{sec5} we give the conclusions and perspectives 
for future work.

\section{Fluctuation operators and dynamic structure factors}
\label{sec2}

\setcounter{equation}{0}

Most fluctuation operators of interest are constructed from local operators
$A_n$ of the model system under consideration:
\begin{eqnarray}
\label{2.01}
A_{\kappa}
=\frac{1}{\sqrt{N}}
\sum_{n=1}^N\exp\left({\mbox{i}}\kappa n\right)A_n.
\end{eqnarray}
Associated with each fluctuation operator (\ref{2.01}) is a dynamic structure factor,
\begin{eqnarray}
\label{2.02}
S_{AA}(\kappa,\omega)
=2\pi\sum_{\lambda\lambda^{\prime}}
\frac{\exp\left(-\beta E_{\lambda^{\prime}}\right)}{Z}
\left\vert
\langle\lambda^{\prime}\vert A_\kappa \vert\lambda\rangle
\right\vert^2
\delta\left(\omega-E_{\lambda}+E_{\lambda^{\prime}}\right),
\end{eqnarray}
which describes fluctuations of a specific kind.
Here $E_\lambda$ and  $\vert\lambda\rangle$
are the eigenvalues and the eigenvectors of $H$,
and $Z$ is the partition function.
Of particular interest is the zero-temperature limit 
$T=0$ (i.e. $\beta\to\infty$),
where the thermal fluctuations fade away
leaving pure quantum fluctuations in the wake.
What remains in (\ref{2.02})
are transitions between the ground state and all excited states
that can be reached by the fluctuation operator $A_\kappa$:
\begin{eqnarray}
\label{2.03}
S_{AA}(\kappa,\omega)
~\stackrel{(\beta\to\infty)}{=}~
2\pi\sum_{\lambda}
\left\vert
\langle {\rm{GS}}\vert A_\kappa \vert \lambda\rangle
\right\vert^2
\delta\left(\omega-\omega_\lambda\right),\quad \omega_\lambda=E_\lambda-E_{\rm{GS}}.
\end{eqnarray}
The dynamically relevant spectrum
observable in (\ref{2.02}) or (\ref{2.03})
may vary considerably between fluctuation operators. Among other things, the
spectrum is sensitive to their symmetry properties.

For the $s=\frac{1}{2}$ transverse $XX$ chain (\ref{1.01}) 
the most important and most widely studied
dynamic structure factors are those for the local spin operators
\begin{equation}
\label{2.04}
s_n^z=c_n^\dagger c_n-\frac{1}{2},
\quad
s_n^+=s_n^x+{\mbox{i}}s_n^y
=c_n^\dagger\exp\left({\mbox{i}}\pi\sum_{j=1}^{n-1}c_j^\dagger c_j\right),
\quad
s_n^-=s_n^x-{\mbox{i}}s_n^y
=\exp\left(-{\mbox{i}}\pi\sum_{j=1}^{n-1}c_j^\dagger c_j\right)c_n.
\end{equation}
At zero temperature
the dynamic spin structure factor $S_{zz}(\kappa,\omega)$ is known
to couple exclusively to the continuum of particle-hole excitations
in the fermion representation,
whereas $S_{xx}(\kappa,\omega)=S_{yy}(\kappa,\omega)$ couples to excitations
involving an arbitrarily high number of fermion excitations from the ground
state \cite{13,14}.

The fluctuation operators considered here
are constructed from local spin operators
on nearest and next-nearest neighbor sites.
The dimer fluctuation operator
$D_\kappa$
and trimer
fluctuation operator
$T_\kappa$
are obtained via (\ref{2.01}) from
\begin{equation}
\label{2.05}
D_n=s_n^xs_{n+1}^x+s_n^ys_{n+1}^y
=\frac{1}{2}\left(c_n^\dagger c_{n+1}-c_nc^\dagger_{n+1}\right)
\end{equation}
and
\begin{equation}
\label{2.06}
T_n=s_n^xs_{n+2}^x+s_n^ys_{n+2}^y
=\frac{1}{2}
\left(
c_n^\dagger c_{n+2}-c_nc^\dagger_{n+2}
-2c_n^\dagger c^\dagger_{n+1}c_{n+1}c_{n+2}
+2c_nc^\dagger_{n+1}c_{n+1}c^\dagger_{n+2}
\right),
\end{equation}
respectively.
There is no unique way of defining dimer and trimer fluctuation operators.
The most suitable choice
depends on the nature and the symmetry of the model system at hand.
The operators (\ref{2.05}) and (\ref{2.06}) have the advantage
that the associated dynamic structure factors
$S_{DD}(\kappa,\omega)$
and
$S_{TT}(\kappa,\omega)$
can be analyzed exactly for the $s=\frac{1}{2}$ transverse $XX$ chain
(\ref{1.01}) in the fermion representation.

As a motivation for the dimer and trimer operators used in this study,
we offer a twofold argument.
For a completely dimerized state,
where {\em nearest-neighbor} spin correlations
alternate between zero and a nonzero value along the chain,
the operator $\sqrt{N}D_\pi$ plays the role of dimer order parameter.
Likewise,
for a completely trimerized state,
where {\em next-nearest neighbor} spin correlations
assume a period-three sequence of values zero, zero, nonzero,
the operator $\sqrt{N}T_{2\pi/3}$ plays the role of trimer order parameter.

Conversely,
if we perturb the uniform $XX$ Hamiltonian (\ref{1.01}) by interactions of the form
\begin{eqnarray}
\label{2.07}
H_D=\varepsilon\sum_{n=1}^N\cos\left(\pi n\right)D_n
\end{eqnarray}
or
\begin{eqnarray}
\label{2.08}
H_T=\varepsilon\sum_{n=1}^N\cos\left(\frac{2\pi}{3} n\right)T_n,
\end{eqnarray}
the ground state becomes dimerized or trimerized, respectively.
In the former case,
nearest neighbor correlations
are modified by period-two perturbative corrections
of order $-\varepsilon, +\varepsilon$
and in the latter case
by period-three corrections
of order $-\frac{1}{2}\varepsilon, -\frac{1}{2}\varepsilon, +\varepsilon$.

We may formally introduce the polymer fluctuation operator of order $l$,
${\cal{P}}_\kappa^{(l)}$, 
via (\ref{2.01}) from
\begin{eqnarray}
\label{2.09}
{\cal{P}}_n^{(l)}
=s_n^xs_{n+l}^x+s_n^ys_{n+l}^y.
\end{eqnarray}
It includes the dimer and trimer operators for $l=1,2$, respectively:
${\cal{P}}_n^{(1)}=D_n$,
${\cal{P}}_n^{(2)}=T_n$.
From the fermion representation of the polymer operator (\ref{2.09}) as carried
out explicitly in (\ref{2.05}) and (\ref{2.06}) for the lowest two orders, it is
evident that the dynamic polymer structure factor $S_\mathcal{PP}(\kappa,\omega)$ at zero temperature
will involve $2m$-fermion excitations with $m=1,2,\ldots,l$ from the ground
state. For an infinitely long chain $(N\to\infty)$ the polymer fluctuation operator and
the function $S_\mathcal{PP}(\kappa,\omega)$ may
thus serve useful roles in attempts to understand the enormously complex
dynamic spin structure factors $S_{xx}(\kappa,\omega)=S_{yy}(\kappa,\omega)$. 
Such tools by which
the complexity of the dynamically relevant excitation spectrum can be gradually
and systematically increased are not only useful for the calculations in the
fermion representation as performed here but also for the recently developed
techniques of calculating transition rates for the $XX$ model in the framework
of the Bethe ansatz \cite{15,16,17}. 
The time-dependent polymer correlation function is related, in the limit $l\to\infty$,
to time-dependent spin correlation functions as follows:
\begin{eqnarray}
\label{2.11}
\langle {\cal{P}}_n^{(l)}(t){\cal{P}}_{n+m}^{(l)}(0) \rangle
~\stackrel{(l\to\infty)}{\to}~
2\langle s^x_n(t)s_{n+m}^x(0)\rangle^2
+
2\langle s^x_n(t)s_{n+m}^y(0)\rangle^2.
\end{eqnarray}
Note that $\langle s^x_n(t)s_{n+m}^y(0)\rangle$ is nonzero only if $\Omega\neq0$.

\section{Two-fermion dynamic structure factors}
\label{sec3}

\setcounter{equation}{0}

We start with the dynamic quantities 
which are governed by particle-hole excitations.
The equilibrium time-dependent correlation functions
for the operators $s_n^z(t)$ and $D_n(t)$
can be evaluated directly:
\begin{eqnarray}
\label{3.01}
\langle s_n^z(t)s_{n+l}^z(0) \rangle -\langle s^z\rangle^2
=\frac{1}{N^2} \sum_{\kappa_1,\kappa_2}
e^{-{\mbox{i}}\left(\kappa_1-\kappa_2\right)l}
\exp\left({\mbox{i}}\left[\Lambda_{\kappa_1}-\Lambda_{\kappa_2}\right]t\right)
n_{\kappa_1}\left(1-n_{\kappa_2}\right),
\end{eqnarray}
\begin{eqnarray}
\label{3.02}
\langle D_n(t)D_{n+l}(0)\rangle -\langle D\rangle^2
=\frac{1}{N^2} \sum_{\kappa_1,\kappa_2} \cos^2\frac{\kappa_1+\kappa_2}{2}
e^{-{\mbox{i}}\left(\kappa_1-\kappa_2\right)l}
\exp\left({\mbox{i}}\left[\Lambda_{\kappa_1}-\Lambda_{\kappa_2}\right]t\right)
n_{\kappa_1}\left(1-n_{\kappa_2}\right),
\end{eqnarray}
where
$n_\kappa=1/\left(1+\exp\left(\beta\Lambda_\kappa\right)\right)$
is the Fermi function
and
\begin{eqnarray}
\label{3.03}
\langle s^z\rangle = \frac{1}{N}\sum_{n=1}^N\langle s^z_n\rangle
=-\frac{1}{2N}\sum_\kappa\tanh\frac{\beta\Lambda_\kappa}{2},
\end{eqnarray}
\begin{eqnarray}
\label{3.04}
\langle D\rangle = \frac{1}{N}\sum_{n=1}^N\langle D_n\rangle
=-\frac{1}{2N}\sum_\kappa\cos\kappa\tanh\frac{\beta\Lambda_\kappa}{2}.
\end{eqnarray}

The associated dynamic structure factors,
\begin{eqnarray}
\label{3.05}
S_{AA}(\kappa,\omega)
=\sum_{l=1}^N\exp\left(-{\mbox{i}}\kappa l\right)
\int_{-\infty}^{\infty}{\mbox{d}}t\exp\left({\mbox{i}}\omega t\right)
\langle \left(A_n(t)-\langle A\rangle\right)
\left(A_{n+l}(0)-\langle A\rangle\right)\rangle,
\end{eqnarray}
all of which involve two-fermion transitions,
are obtained by Fourier transform.
The resulting expressions for
$N\to\infty$
can be brought into the form
\begin{eqnarray}
\label{3.06}
S_{zz}(\kappa,\omega) 
=\int_{-\pi}^\pi{\mbox{d}}\kappa_1n_{\kappa_1}\left(1-n_{\kappa_1+\kappa}\right)
\delta\left(\omega+\Lambda_{\kappa_1}-\Lambda_{\kappa_1+\kappa}\right)
=\sum_{\kappa^\star} \frac{n_{\kappa^\star}\left(1-n_{\kappa+\kappa^\star}\right)}
{2\vert J\sin\frac{\kappa}{2}\cos\left(\frac{\kappa}{2}+\kappa^\star\right)\vert},
\end{eqnarray}
\begin{eqnarray}
\label{3.07}
S_{DD}(\kappa,\omega) 
=
\sum_{\kappa^\star} \frac{\cos^2\left(\frac{\kappa}{2}+\kappa^\star\right)
n_{\kappa^\star}\left(1-n_{\kappa+\kappa^\star}\right)}
{2\vert J\sin\frac{\kappa}{2}\cos\left(\frac{\kappa}{2}+\kappa^\star\right)\vert},
\end{eqnarray}
where $-\pi\leq\kappa^\star\leq\pi$ are the solutions of the equation
\begin{eqnarray}
\label{3.08}
\omega=-2J\sin\frac{\kappa}{2}\sin\left(\frac{\kappa}{2}+\kappa^\star\right).
\end{eqnarray}

The dynamic structure factors (\ref{3.06}), (\ref{3.07})
are governed by the two-fermion (particle-hole) excitation continuum
the properties of which were examined in Refs. \cite{18,19}.
This continuum is well visible in Figs. \ref{fig01} and \ref{fig02} below,
which we include mainly for comparison with new results.
At zero temperature, $T=0$,
the two-fermion excitation continuum
has the following lower, middle and upper boundaries 
in the $(\kappa,\omega)$-plane\footnote
{We assume that $0\leq\kappa\leq\pi$ in the rest equations of this Section;   
these equations are valid also for $-\pi\leq\kappa\leq 0$
after the change $\kappa\to -\kappa$.}
\begin{eqnarray}
\label{3.09}
\frac{\omega_l}{\vert J\vert}
=2\sin\frac{\kappa}{2}
\left\vert\sin\left(\frac{\kappa}{2}-\alpha\right)\right\vert,
\end{eqnarray}
\begin{eqnarray}
\label{3.10}
\frac{\omega_m}{\vert J\vert}
=2\sin\frac{\kappa}{2}\sin\left(\frac{\kappa}{2}+\alpha\right),
\end{eqnarray}
\begin{eqnarray}
\label{3.11}
\frac{\omega_u}{\vert J\vert}
=
\left\{
\begin{array}{ll}
{\displaystyle 2\sin\frac{\kappa}{2}\sin\left(\frac{\kappa}{2}+\alpha\right)}, &
{\mbox{if}}\;\;\;0\leq \kappa\leq\pi-2\alpha, \\
{\displaystyle 2\sin\frac{\kappa}{2}}, &
{\mbox{if}}\;\;\;\pi-2\alpha\leq\kappa\leq\pi,
\end{array}
\right.
\end{eqnarray}
respectively. The parameter $\alpha=\arccos\left(\Omega/\vert J\vert\right)$
varies from $\pi$ when $\Omega=-\vert J\vert$
to $0$ when $\Omega=\vert J\vert$.
The $\omega$-profiles at fixed $\kappa$ 
of the two-fermion dynamic structure factors
may exhibit square-root divergences
(a common density-of-states effect
in one dimension)
when $\omega\to2|J|\sin(\kappa/2)$.
At $T>0$ the lower boundary of two-fermion excitation continuum is smeared
out. The spectral weight in (\ref{3.06}) and (\ref{3.07}) is now confined to
$|\omega|\leq2|J|\sin(\kappa/2)$.

Closed-form expressions 
for the two-fermion dynamic structure factors (\ref{3.06}), (\ref{3.07})
exist in the low-temperature and high-temperature limits. At $T=0$ we have 
\begin{eqnarray}
\label{3.13}
S_{zz}(\kappa,\omega) 
=
\frac{1}{\sqrt{4J^2\sin^2\frac{\kappa}{2}-\omega^2}}
\cdot
\left\{
\begin{array}{ll}
\Theta(\omega-\omega_l)\Theta(\omega_u-\omega), & 
{\mbox{if}}\;\;\; 0\leq\kappa\leq\pi- 2\alpha,\\
\left(\Theta(\omega-\omega_l)+\Theta(\omega-\omega_m)\right)\Theta(\omega_u-\omega), & 
{\mbox{if}}\;\;\; \pi- 2\alpha\leq\kappa\leq\pi,
\end{array}
\right.
\end{eqnarray}
\begin{eqnarray}
\label{3.14}
S_{DD}(\kappa,\omega) 
=
\frac{ \sqrt{ 4J^2\sin^2\frac{\kappa}{2}-\omega^2}}{ 4J^2\sin^2\frac{\kappa}{2}}
\cdot
\left\{
\begin{array}{ll}
\Theta(\omega-\omega_l)\Theta(\omega_u-\omega), & 
{\mbox{if}}\;\;\; 0\leq\kappa\leq\pi- 2\alpha,\\
\left(\Theta(\omega-\omega_l)+\Theta(\omega-\omega_m)\right)\Theta(\omega_u-\omega), & 
{\mbox{if}}\;\;\; \pi- 2\alpha\leq\kappa\leq\pi,
\end{array}
\right.
\end{eqnarray}
and at $T\to\infty$ we have
\begin{eqnarray}
\label{3.15}
S_{zz}(\kappa,\omega)
=
\frac{1}{2\sqrt{4J^2\sin^2\frac{\kappa}{2}-\omega^2}}
\Theta\left(2\vert J\vert \sin\frac{\kappa}{2}-|\omega|\right),
\end{eqnarray}
\begin{eqnarray}
\label{3.16}
S_{DD}(\kappa,\omega)
=
\frac{\sqrt{4J^2\sin^2\frac{\kappa}{2}-\omega^2}}{8J^2\sin^2\frac{\kappa}{2}}
\Theta\left(2\vert J\vert \sin\frac{\kappa}{2}-|\omega|\right).
\end{eqnarray}
The zero-temperature results 
for $S_{zz}(\kappa,\omega)$ can be found in Eq.~(2.3) of Ref.~\cite{18} 
and 
for $S_{DD}(\kappa,\omega)$ at $\Omega=0$ in Eq.~(3.2) of Ref.~\cite{08}.

In Figs.  \ref{fig01} and \ref{fig02} 
we show the dynamic structure factors (\ref{3.06}), (\ref{3.07})
at zero temperature $T=0$ and different values of the transverse field (panels a,
b, c), and at $T\to\infty$ (panel d). 
The results for $T\to\infty$ are independent of $\Omega$. 
As we can see, 
the two-fermion dynamic structure factors 
are nonzero within the two-fermion excitation continuum
in the $(\kappa,\omega)$-plane.
Their spectral-weight distributions 
are controlled 
by the Fermi functions, 
the multiplicity of the solution of Eq.~(\ref{3.08}),
the singularities in the density of one-particle states, 
and the explicit form of the rest of the integrand in (\ref{3.06}), (\ref{3.07}).
Another two-fermion dynamic quantity will be presented in Sec.~\ref{sec4}, namely
the two-fermion contribution to the dynamic trimer structure factor,
$S_{TT}^{(2)}(\kappa,\omega)$.

\section{Four-fermion dynamic structure factor}
\label{sec4}

\setcounter{equation}{0}

Next we consider the dynamics of the trimer fluctuations.  The method remains
the same but its execution is more tedious.  In addition to two-fermion
transitions also four-fermion transitions contribute to the trimer fluctuations.
The expression for the equilibrium time-dependent trimer-trimer correlation function reads
\begin{eqnarray}
\label{4.01}
\langle T_n(t) T_{n+l}(0)\rangle - \langle T\rangle^2
=
\frac{1}{N^2}\sum_{k_1,k_2}
C^{(2)}\left(\kappa_1,\kappa_2\right)
e^{-{\mbox{i}}\left(\kappa_1-\kappa_2\right)l}
\exp\left({\mbox{i}}\left[\Lambda_{\kappa_1}-\Lambda_{\kappa_2}\right]t\right)
n_{\kappa_1}\left(1-n_{\kappa_2}\right)
\nonumber\\
+
\frac{1}{N^4}\sum_{\kappa_1,\kappa_2,\kappa_3,\kappa_4}
C^{(4)}\left(\kappa_1,\kappa_2,\kappa_3,\kappa_4\right)
e^{-{\mbox{i}}\left(\kappa_1+\kappa_2-\kappa_3-\kappa_4\right)l}
\exp\left({\mbox{i}}\left[\Lambda_{\kappa_1}+\Lambda_{\kappa_2}
-\Lambda_{\kappa_3}-\Lambda_{\kappa_4}\right]t\right)
\nonumber\\
\times~ 
n_{\kappa_1}n_{\kappa_2}\left(1-n_{\kappa_3}\right)\left(1-n_{\kappa_4}\right),
\end{eqnarray}
where
\begin{eqnarray}
\label{4.02}
\langle T\rangle=\frac{1}{N}\sum_{n=1}^N\langle T_n\rangle 
=c_2+2c_1^2-2c_0c_2,
\end{eqnarray}
\begin{eqnarray}
\label{4.03}
C^{(2)}\left(\kappa_1,\kappa_2\right)
=
\left(1-2c_0\right)^2\cos^2\left(\kappa_1+\kappa_2\right)
+4c_1\left(1-2c_0\right)
\left(\cos^2\left(\kappa_1+\frac{\kappa_2}{2}\right)
+\cos^2\left(\frac{\kappa_1}{2}+\kappa_2\right)\right)
\nonumber\\
+4c_1^2\left(\cos^2\kappa_1+\cos^2\kappa_2\right)
+8\left(-c_2+c_1^2+2c_0c_2\right)
\cos^2\frac{\kappa_1+\kappa_2}{2}
+8c_1^2\cos^2\frac{\kappa_1-\kappa_2}{2}
\nonumber\\
+4c_1\left(1-2c_0-4c_2\right)
\left(\cos^2\frac{\kappa_1}{2}+\cos^2\frac{\kappa_2}{2}\right)
\nonumber\\
+4c_2-8c_1-8c_1^2+4c_2^2+ 16c_0c_1-8c_0c_2+16c_1c_2,
\end{eqnarray}
\begin{eqnarray}
\label{4.04}
C^{(4)}\left(\kappa_1,\kappa_2,\kappa_3,\kappa_4\right)
=16
\sin^2\frac{\kappa_1-\kappa_2}{2}
\sin^2\frac{\kappa_3-\kappa_4}{2}
\cos^2\frac{\kappa_1+\kappa_2+\kappa_3+\kappa_4}{2}.
\end{eqnarray}
Here we have introduced the function 
$c_p=\left(1/N\right)\sum_\kappa\cos\left(p\kappa\right)n_\kappa$.
For $N\to\infty$ and
at zero temperature we have
$c_0=1$ if $\Omega\leq -\vert J\vert$,
$c_0=\alpha/\pi$ if $-\vert J\vert\leq\Omega\leq\vert J\vert$,
$c_0=0$ if $\vert J\vert\leq\Omega$,
$c_p=\left(-{\mbox{sgn}}(J)\right)^p\sin\left(p\alpha\right)/\left(p\pi\right)$
if $\vert\Omega\vert\leq\vert J\vert$ 
and 
$c_p=0$ otherwise ($p=1,2,\ldots$ ).
At $T\to\infty$ we have 
$c_p=\frac{1}{2}\delta_{p,0}$.
The resulting dynamic trimer structure factor (\ref{3.05})
for $N\to\infty$ then has the following form:
\begin{eqnarray}
\label{4.05}
S_{TT}(\kappa,\omega)
=S^{(2)}_{TT}(\kappa,\omega)
+S^{(4)}_{TT}(\kappa,\omega),
\end{eqnarray}
where
\begin{equation}
  \label{eq:1}
 S^{(2)}_{TT}(\kappa,\omega)
=
\int_{-\pi}^\pi{\mbox{d}}\kappa_1
C^{(2)}\left(\kappa_1,\kappa_1+\kappa\right)
n_{\kappa_1}\left(1-n_{\kappa_1+\kappa}\right)
\delta\left(\omega+\Lambda_{\kappa_1}-\Lambda_{\kappa_1+\kappa}\right), 
\end{equation}
\begin{eqnarray}
  \label{eq:2}
 S^{(4)}_{TT}(\kappa,\omega)
&=& \frac{1}{4\pi^2}
\int_{-\pi}^\pi{\mbox{d}}\kappa_1\int_{-\pi}^\pi{\mbox{d}}\kappa_2
\int{\mbox{d}}\kappa_3
C^{(4)}\left(\kappa_1,\kappa_2,\kappa_3,\kappa_1+\kappa_2-\kappa_3+\kappa\right)
\nonumber \\
&\times&
n_{\kappa_1}n_{\kappa_2}
\left(1-n_{\kappa_3}\right)\left(1-n_{\kappa_1+\kappa_2-\kappa_3+\kappa}\right)
\delta\left(\omega+\Lambda_{\kappa_1}+\Lambda_{\kappa_2}
-\Lambda_{\kappa_3}-\Lambda_{\kappa_1+\kappa_2-\kappa_3+\kappa}\right). 
\end{eqnarray}

The spectral weight in this quantity comes from both the two-fermion (one
particle and one hole) excitation continuum and the four-fermion (two particles
and two holes) excitation continuum.  Let us first discuss the properties of the
four-fermion excitation continuum and then the properties of $S_{TT}(\kappa,\omega)$.  At
$T=0$ the four-fermion excitation continuum (for $J=-\vert J\vert<0$) is
determined by the conditions

\begin{eqnarray}
\label{4.06}
\frac{\omega}{\vert J\vert}
=\cos\kappa_1+\cos\kappa_2-\cos\kappa_3-\cos\kappa_4,
\;\;\;
\kappa=-\kappa_1-\kappa_2+\kappa_3+\kappa_4\;({\mbox{mod}}(2\pi)),
\nonumber\\
\cos\kappa_1 \geq \frac{\Omega}{\vert J\vert},\;\;\;
\cos\kappa_2 \geq \frac{\Omega}{\vert J\vert},\;\;\;
\cos\kappa_3 \leq \frac{\Omega}{\vert J\vert},\;\;\;
\cos\kappa_4 \leq \frac{\Omega}{\vert J\vert},
\end{eqnarray}
$-\pi\leq\kappa_{1,2,3}\leq\pi$,
$-\pi\leq\kappa\leq\pi$.
Equations (\ref{4.06}) imply
that the four-fermion excitation continuum 
(like the two-fermion excitation continuum)
exists only if the magnetic field does not exceed the saturation field:
$\vert\Omega\vert\leq\vert J\vert$.

An analytical expression
for the lower boundary of the four-fermion excitation continuum
in the $(\kappa,\omega)$-plane 
depends on $\Omega$ and $\kappa$
and is given by one of the following expressions
(see Appendix for additional details):
\begin{eqnarray}
\label{4.07}
\frac{\omega_l^{(1)}}{\vert J\vert}
=2\sin\frac{\vert\kappa\vert}{2}\sin\left(\alpha-\frac{\vert\kappa\vert}{2}\right),
\end{eqnarray}
\begin{eqnarray}
\label{4.08}
\frac{\omega_l^{(2)}}{\vert J\vert}
=4\cos\frac{\kappa}{4}\cos\left(\alpha+\frac{\vert\kappa\vert}{4}\right),
\end{eqnarray}
\begin{eqnarray}
\label{4.09}
\frac{\omega_l^{(3)}}{\vert J\vert}
=-2\sin\left(\alpha+\frac{\vert\kappa\vert}{2}\right)
\sin\left(2\alpha+\frac{\vert\kappa\vert}{2}\right),
\end{eqnarray}
\begin{eqnarray}
\label{4.10}
\frac{\omega_l^{(4)}}{\vert J\vert}
=-2\sin\left(\alpha-\frac{\vert\kappa\vert}{2}\right)
\sin\left(2\alpha-\frac{\vert\kappa\vert}{2}\right),
\end{eqnarray}
\begin{eqnarray}
\label{4.11}
\frac{\omega_l^{(5)}}{\vert J\vert}
=-4\sin\frac{\vert\kappa\vert}{4}\sin\left(\alpha-\frac{\vert\kappa\vert}{4}\right).
\end{eqnarray}
The range in $\kappa$ over which for a given $\Omega$
one of the expressions (\ref{4.07}) -- (\ref{4.11})
forms the lower boundary of the four-fermion continuum
can be read off Fig. \ref{fig03}.
The darkness in this gray-scale plot is a measure of the size of the energy
threshold of the four-fermion continuum (white means zero excitation energy i.e. a
soft mode).  
The boundary between the region $i$ (where $\omega^{(i)}_l$ is the lower
boundary) and the region $j$ (where $\omega^{(j)}_l$ is the lower boundary) follows
from the matching condition $\omega^{(i)}_l=\omega^{(j)}_l$ and is given by the formula
$\vert\kappa\vert=l_{ij}(\alpha)$ where
\begin{eqnarray}
\label{4.12}
l_{12}(\alpha)=4\arctan\frac{\tan\alpha-\sqrt{\tan^2\alpha-3}}{3},
\;\;\;
\vert\kappa\vert\leq \frac{2\pi}{3},
\end{eqnarray}
\begin{eqnarray}
\label{4.13}
l_{13}(\alpha)=\pi-\alpha,
\;\;\;
\frac{\pi}{2}\leq\vert\kappa\vert\leq \frac{2\pi}{3},
\end{eqnarray}
\begin{eqnarray}
\label{4.14}
l_{14}(\alpha)=2\alpha,
\end{eqnarray}
\begin{eqnarray}
\label{4.15}
l_{23}(\alpha)=2\pi-4\alpha,
\end{eqnarray}
\begin{eqnarray}
\label{4.16}
l_{34}(\alpha)=\vert\kappa\vert+\cos\alpha-\frac{1}{2},
\;\;\;
\frac{2\pi}{3}\leq\vert\kappa\vert\leq \pi,
\end{eqnarray}
\begin{eqnarray}
\label{4.17}
l_{45}(\alpha)=4\alpha.
\end{eqnarray}
The boundary between regions 2 and 4 is determined by the
cubic equation
\begin{eqnarray}
\label{4.18}
\left(\sin\vert\kappa\vert-2\sin\frac{\vert\kappa\vert}{2}\right)\tan^3\alpha
+
\left(3+2\cos\frac{\kappa}{2}+3\cos\kappa\right)\tan^2\alpha
\nonumber\\
-
\left(2\sin\frac{\vert\kappa\vert}{2}+3\sin\vert\kappa\vert\right)\tan\alpha
+
3+2\cos\frac{\kappa}{2}-\cos\kappa=0.
\end{eqnarray}
Typical lower boundaries of the four-fermion continuum for several values of $\Omega$
can be seen in Fig. \ref{fig04}.
The soft modes according to (\ref{4.07}) -- (\ref{4.11})
are given by
\begin{eqnarray}
\label{4.19}
\vert\kappa_0\vert = \left\{0, 2\pi-4\alpha, 2\alpha, 4\alpha\right\}.
\end{eqnarray}
Alternatively,
the soft modes (\ref{4.19})
may be determined directly from (\ref{4.06}).
They occur when
$\cos\kappa_1=\cos\kappa_2=\cos\kappa_3=\cos\kappa_4=\cos\alpha$.

The upper boundary of the four-fermion continuum for
$0\leq\Omega\leq\vert J\vert/\sqrt{2}$
is given by
\begin{eqnarray}
\label{4.20}
\frac{\omega_u^{(1)}}{\vert J\vert}
=4\cos\frac{\kappa}{4}.
\end{eqnarray}
For
$\vert J\vert/\sqrt{2}\leq\Omega\leq\vert J\vert$
the upper boundary is given by (\ref{4.20})
only if
$\vert\kappa\vert\leq 4\alpha$
whereas, if
$4\alpha\leq\vert\kappa\vert\leq \pi$,
it is given by
\begin{eqnarray}
\label{4.21}
\frac{\omega_u^{(2)}}{\vert J\vert}
=4\cos\frac{\kappa}{4}\cos\left(\alpha-\frac{\vert\kappa\vert}{4}\right)
\end{eqnarray}
(see Appendix).
The upper boundaries of the four-fermion continuum for several values of $\Omega$
can be seen in Fig. \ref{fig04} in comparison with the corresponding two-fermion
continuum. 

The four-fermion continuum always contains the two-fermion continuum.
The lower boundaries coincide in part. The upper boundaries are different.
In the zero field case we have 
$\omega_l=\vert J\vert\sin\vert\kappa\vert$ for both continua.
The upper boundaries are  
$\omega_u=4\vert J\vert\cos\frac{\kappa}{4}$
and  
$\omega_u=2\vert J\vert\sin\frac{\vert\kappa\vert}{2}$
for the two-fermion and four-fermion continua, respectively.
As the saturation field $\Omega=|J|$ is approached from below, 
the two-fermion continuum narrows to a branch 
and then disappears whereas the four-fermion continuum remains an extended
region, bounded by 
$\omega_l=4\vert J\vert\sin^2\frac{\kappa}{4}$
and
$\omega_u=4\vert J\vert\cos^2\frac{\kappa}{4}$, 
and then disappears abruptly.

Now consider the equation
\begin{eqnarray}
\label{4.22}
\sqrt{\sum_{j=1}^3\left(\frac{\partial}{\partial\kappa_j}
\left[\cos\kappa_1+\cos\kappa_2
-\cos\kappa_3-\cos\left(\kappa+\kappa_1+\kappa_2-\kappa_3\right)\right]\right)^2} = 0.
\end{eqnarray}
It is satisfied for
\begin{eqnarray}
\label{4.23}
\frac{\omega^{(1)}_s}{\vert J\vert}
=2\sin\frac{\vert\kappa\vert}{2},
\end{eqnarray}
\begin{eqnarray}
\label{4.24}
\frac{\omega^{(2)}_s}{\vert J\vert}
=4\sin\frac{\vert\kappa\vert}{4},
\end{eqnarray}
\begin{eqnarray}
\label{4.25}
\frac{\omega^{(3)}_s}{\vert J\vert}
=4\cos\frac{\kappa}{4}.
\end{eqnarray}
Thus,
for $\kappa$ or $\omega$ approaching the curves (\ref{4.23}) -- (\ref{4.25})
in $(\kappa,\omega)$-space, the quantity
\begin{eqnarray}
\label{4.26}
S(\kappa,\omega)
&=& \int_{-\pi}^{\pi}{\mbox{d}}\kappa_1
\int_{-\pi}^{\pi}{\mbox{d}}\kappa_2
\int_{-\pi}^{\pi}{\mbox{d}}\kappa_3
S\left(\kappa_1,\kappa_2,\kappa_3,\kappa\right)
\nonumber\\
&\times&
\delta\left(\omega-\vert J\vert\cos\kappa_1-\vert J\vert\cos\kappa_2
+\vert J\vert\cos\kappa_3+\vert J\vert\cos\left(\kappa+\kappa_1+\kappa_2-\kappa_3\right)\right)
\end{eqnarray}
exhibits cusp singularities
(akin to density-of-states effects in three dimensions).
The exact nature of the cusps also depends on the factor
$S\left(\kappa_1,\kappa_2,\kappa_3,\kappa\right)$, which varies between different dynamic structure
factors with a four-fermion part. It always includes the factor 
$n_{\kappa_1}n_{\kappa_2}
\left(1-n_{\kappa_3}\right)\left(1-n_{\kappa_1+\kappa_2-\kappa_3+\kappa}\right)$ as can be seen in
expression (\ref{eq:2}). 
In Fig. \ref{fig05}
we show the $\omega$-dependence of $S(\kappa,\omega)$ as given by (\ref{4.26})
at $\kappa=0,\;\frac{\pi}{2},\;\frac{2\pi}{3},\;\pi$
when $S\left(\kappa_1,\kappa_2,\kappa_3,\kappa\right)=1$
and when
$S\left(\kappa_1,\kappa_2,\kappa_3,\kappa\right)
=
n_{\kappa_1}n_{\kappa_2}
\left(1-n_{\kappa_3}\right)\left(1-n_{\kappa_1+\kappa_2-\kappa_3+\kappa}\right)$
for several values of $\Omega$.

At $T>0$ the lower boundary of the four-fermion excitation continuum is smeared out
and the upper boundary becomes $\omega_u=4\vert J\vert\cos\frac{\kappa}{4}$.

The properties of the multimagnon continua
of quantum spin chains have been examined in some detail 
in the recent paper of Barnes \cite{20}.
In particular,
the lower/upper boundary of the two- and higher magnon continua
were determined.
It was shown that the boundary curves
under certain conditions
may exhibit discontinuous changes in composition and cusps.
Moreover,
a behavior of the density of (two- and higher magnon) states
on the continuum boundaries and within the continuum
was considered
and the existence of discontinuities was pointed out.
These features of one-dimensional quantum spin systems
are expected to become accessible experimentally
in high-resolution inelastic neutron scattering
on alternating chain and ladder materials.

Interestingly,
the $s=\frac{1}{2}$ transverse $XX$ chain
which can be mapped onto noninteracting spinless fermions
provides an excellent example of a system
whose dynamic properties are governed by continua of multifermion excitations.
In particular,
the dynamics of trimer fluctuations 
provides a direct motivation for analyzing the four-fermion excitation continuum.
Unlike in the analysis reported in Ref. \cite{20},
where the statistics of the elementary excitations (magnons)
is not known, here the quasiparticles are known to be fermions and the
consequences are fully taken into account.

Finally,
let us examine the explicit expression
for the dynamic trimer structure factor $S_{TT}(\kappa,\omega)$ (\ref{4.05}).
In Figs. \ref{fig06},  \ref{fig07},  \ref{fig08},  \ref{fig09} 
we present the zero-temperature dynamic trimer structure factor
at different values of $\Omega$.
In Fig.  \ref{fig10} 
we present the same quantity at infinite temperature.
We show separately 
the two-fermion contribution (panels a) 
and the four-fermion contribution (panels b)  
as well as the sum of these contributions (panels c).
We observe how the spectral weight is spread across the four-fermion continuum. 
We also see that the two-fermion contribution stands out in terms of spectral weight.
The two-fermion and four-fermion contributions 
are comparable in intensity at $T=0$ and small $\Omega$.
As $\Omega$ increases the two-fermion contribution becomes more important 
and it completely dominates as $\Omega\to\vert J\vert$.
In the high-temperature limit 
the two-fermion contribution is very dominant but the four-fermion continuum is
still in evidence.

\section{Conclusions}
\label{sec5}

\setcounter{equation}{0}

In summary,
we have investigated some aspects of the dynamics 
of the $s=\frac{1}{2}$ transverse $XX$ chain
examining, 
in particular,
the dynamics of dimer and trimer operators.
For this purpose we have calculated several dynamic structure factors 
on a rigorous basis 
within the Jordan-Wigner representation.
While the dynamic dimer structure factor $S_{DD}(\kappa,\omega)$
and the dynamic transverse spin structure factor $S_{zz}(\kappa,\omega)$ 
are governed by fermionic one-particle-one-hole excitations,
the dynamic trimer structure factor $S_{TT}(\kappa,\omega)$
also contains contributions from two-particle-two-hole excitations. 
We have described the structure of the two-fermion and four-fermion excitation
spectra in some 
detail and investigated the distribution of spectral weight in $S_{TT}(\kappa,\omega)$ across these
continua at zero and nonzero temperature. 
In particular,
we have established
the boundaries of the four-fermion spectral range,
the locations of soft modes,
and the singularity structure, 
which includes one-dimensional 
and
three-dimensional 
density-of-states
features 
(van Hove singularities).

An alternative technique
to evaluate dynamic structure factor of quantum spin chains 
is based on the Bethe ansatz solutions \cite{05,16,17}.
Recently such an approach 
has been applied to the $s=\frac{1}{2}$ $XX$ chain \cite{16}.
Moreover, the relation between spinons or magnon quasiparticles
and Jordan-Wigner fermions was discussed in some detail.
It will be interesting 
to interpret the two-fermion and four-fermion excitations 
discussed here
in terms of the Bethe ansatz solution 
as studied in Ref. \cite{16}.

\section*{Acknowledgments}

This study was performed within the framework of the STCU project \#1673.
O. D., J. S. and G. M.
thank
the Wilhelm und Else Heraeus-Stiftung
for the kind hospitality
during the 288. WE-Heraeus-Seminar on the Theme of
``Quantum Magnetism: Microscopic Techniques
for Novel States of Matter''
(Bad Honnef, 2002)
when the present study was launched.
O. D. acknowledges the kind hospitality of the University of Dortmund 
in the spring of 2003
when part of the work was done.
O. D. expresses gratitude
to the Max-Planck-Institut f\"{u}r Physik komplexer Systeme (Dresden)
for its hospitality in the spring of 2004.

\section*{Appendix}
\label{seca}

\renewcommand{\theequation}{A.\arabic{equation}}
\setcounter{equation}{0}

To find the lower and upper boundaries of the four-fermion excitation continuum (\ref{4.06})
at fixed $\Omega$
and $-\pi\leq\kappa\leq\pi$
we search for the extrema of $\omega/\vert J\vert$ as given in (\ref{4.06}) and the 
values of
$\kappa_1$,
$\kappa_2$,
$\kappa_3$,
$\kappa_4$
at which such extrema occur.
Typical results are reported in Fig. \ref{fig11} (lower boundary) and
Fig. \ref{fig12} (upper boundary).

In Fig. \ref{fig11} we show the dependence on $\vert\kappa\vert$  of $\kappa_1$, $\kappa_2$,
$\kappa_3$, $\kappa_4$ where $\omega/\vert J\vert$ assumes a local minimum. We distinguish
five different regions. The global minimum yields the lower continuum boundary. 
If $0\leq\kappa\leq\kappa_{a}$
\begin{eqnarray}
\label{a.01}
\kappa_1=\alpha-\kappa,
\;\;\;
\kappa_2=\kappa_3=\kappa_4=\alpha
\end{eqnarray}
and
\begin{eqnarray}
\label{a.02}
\frac{\omega_l}{\vert J\vert}
=\cos\left(\alpha-\kappa\right)-\cos\alpha
=2\sin\frac{\kappa}{2}\sin\left(\alpha-\frac{\kappa}{2}\right)
=\frac{\omega_l^{(1)}}{\vert J\vert};
\end{eqnarray}
if $\kappa_a\leq\kappa\leq\kappa_b$
\begin{eqnarray}
\label{a.03}
\kappa_1=\kappa_2=\alpha,
\;\;\;
\kappa_3=\kappa_4=\frac{\kappa}{2}+\alpha-\pi
\end{eqnarray}
and
\begin{eqnarray}
\label{a.04}
\frac{\omega_l}{\vert J\vert}
=2\cos\alpha -2\cos\left(\frac{\kappa}{2}+\alpha-\pi\right)
=4\cos\frac{\kappa}{4}\cos\left(\alpha+\frac{\kappa}{4}\right)
=\frac{\omega_l^{(2)}}{\vert J\vert}
\end{eqnarray}
etc.
The values of $\kappa_a$, $\kappa_b$, $\kappa_c$, $\kappa_d$ follow from the matching conditions.

In Fig. \ref{fig12} we show the dependence on $\vert\kappa\vert$  of $\kappa_1$, $\kappa_2$,
$\kappa_3$, $\kappa_4$ where $\omega/\vert J\vert$ assumes a local maximum. We distinguish
two different regions. 
The global maximum yields the upper continuum boundary.
If $0\leq\kappa\leq\kappa_{A}$
\begin{eqnarray}
\label{a.05}
\kappa_1=\kappa_2=-\frac{\kappa}{4},
\;\;\;
\kappa_3=\kappa_4=-\pi+\frac{\kappa}{4}
\end{eqnarray}
and
\begin{eqnarray}
\label{a.06}
\frac{\omega_u}{\vert J\vert}
=4\cos\frac{\kappa}{4}
=\frac{\omega_u^{(1)}}{\vert J\vert};
\end{eqnarray}
if $\kappa_{A}\leq\kappa\leq \pi$
\begin{eqnarray}
\label{a.07}
\kappa_1=\kappa_2=-\alpha,
\;\;\;
\kappa_3=\kappa_4=-\alpha-\pi+\frac{\kappa}{2}
\end{eqnarray}
and
\begin{eqnarray}
\label{a.08}
\frac{\omega_u}{\vert J\vert}
=2\cos\alpha+2\cos\left(\alpha-\frac{\kappa}{2}\right)
=4\cos\frac{\kappa}{4}\cos\left(\alpha-\frac{\kappa}{4}\right)
=\frac{\omega_u^{(2)}}{\vert J\vert}
\end{eqnarray}
etc.
From the matching condition we find $\kappa_{A}=4\alpha$.

\clearpage

\renewcommand\baselinestretch {1.3}
\large\normalsize

FIGURE CAPTIONS

\vspace{5mm}

Fig. 1.
$S_{zz}(\kappa,\omega)$  
at $T=0$
and 
(a) $\Omega=0$,
(b) $\Omega=0.3$,
(c) $\Omega=0.9$,
and (d) at $T\to\infty$ (independent of $\Omega$; only $\omega\ge 0$ is shown).

\vspace{5mm}

Fig. 2.
$S_{DD}(\kappa,\omega)$ 
at $T=0$
and 
(a) $\Omega=0$,
(b) $\Omega=0.3$,
(c) $\Omega=0.9$,
and (d) at $T\to\infty$ (independent of $\Omega$; only $\omega\ge 0$ is shown).

\vspace{5mm}

Fig. 3.
Lower boundary $\omega_l=\mathrm{min}\left(\omega_l^{(j)}\right)$, $j=1,\ldots,5$ of the
four-fermion excitation continuum versus wave number $\kappa$ and transverse field
$\Omega$ (for $\vert J\vert=1$).

\vspace{5mm}

Fig. 4.
Lower and upper boundaries
of the two-fermion and four-fermion continua
for $\vert J\vert=1$
and
$\Omega=0$,
$\Omega=0.3$,
$\Omega=0.6$,
$\Omega=0.9$.
The two-fermion continuum is shown shaded.

\vspace{5mm}

Fig. 5.
$S(\kappa,\omega)$ as given in (\ref{4.26}) versus $\omega$
at $\kappa=0,\;\frac{\pi}{2},\;\frac{2\pi}{3},\;\pi$
with
$S\left(\kappa_1,\kappa_2,\kappa_3,\kappa\right)=1$ (bold curves),
and
$S\left(\kappa_1,\kappa_2,\kappa_3,\kappa\right)=n_{\kappa_1}n_{\kappa_2}
\left(1-n_{\kappa_3}\right)\left(1-n_{\kappa_1+\kappa_2-\kappa_3+\kappa}\right)$
for
$\Omega=0$ (solid curves),
$\Omega=0.3$ (long-dashed curves),
$\Omega=0.6$ (short-dashed curves),
$\Omega=0.9$ (dotted curves).
Vertical lines mark the values of $\omega_s^{(j)}$,
$j=1,2,3$ as given in (\ref{4.23}) -- (\ref{4.25}). Note the different vertical
scales left and right.

\vspace{5mm}

Fig. 6.
$S_{TT}(\kappa,\omega)$
at $T=0$
and 
$\Omega=0$.
Separate plots are shown for  
(a) $S_{TT}^{(2)}(\kappa,\omega)$,
(b) $S_{TT}^{(4)}(\kappa,\omega)$, and 
(c) the sum $S_{TT}(\kappa,\omega)$.

\vspace{5mm}

Fig. 7.
$S_{TT}(\kappa,\omega)$
at $T=0$
and 
$\Omega=0.3$.
Separate plots are shown for  
(a) $S_{TT}^{(2)}(\kappa,\omega)$,
(b) $S_{TT}^{(4)}(\kappa,\omega)$, and 
(c) the sum $S_{TT}(\kappa,\omega)$.

\vspace{5mm}

Fig. 8.
$S_{TT}(\kappa,\omega)$
at $T=0$
and 
$\Omega=0.6$.
Separate plots are shown for  
(a) $S_{TT}^{(2)}(\kappa,\omega)$,
(b) $S_{TT}^{(4)}(\kappa,\omega)$, and 
(c) the sum $S_{TT}(\kappa,\omega)$.

\vspace{5mm}

Fig. 9.
$S_{TT}(\kappa,\omega)$
at $T=0$
and 
$\Omega=0.9$.
Separate plots are shown for  
(a) $S_{TT}^{(2)}(\kappa,\omega)$,
(b) $S_{TT}^{(4)}(\kappa,\omega)$, and 
(c) the sum $S_{TT}(\kappa,\omega)$.

\vspace{5mm}

Fig. 10.
$S_{TT}(\kappa,\omega)$
at $T\to\infty$
(independent of $\Omega$; only $\omega\ge 0$ is shown).
Separate plots are shown for  
(a) $S_{TT}^{(2)}(\kappa,\omega)$,
(b) $S_{TT}^{(4)}(\kappa,\omega)$, and 
(c) the sum $S_{TT}(\kappa,\omega)$.

\vspace{5mm}

Fig. 11.
Search for the lower boundary of the four-fermion excitation continuum.
Shown are the values of $\kappa_1$, $\kappa_2$, $\kappa_3$, $\kappa_4$
at which a minimum of $\omega/\vert J\vert$ as given in (\ref{4.06}) occurs
at $\Omega=0.3\vert J\vert$
and $-\pi\leq\kappa\leq\pi$.
The dependences $\kappa_1$ and $\kappa_3$ on $\kappa$ are shown by dashed curves,
the dependences $\kappa_2$ and $\kappa_4$ on $\kappa$ are shown by dotted curves,
the dependence of the minimal value of $\omega/\vert J\vert$ on $\kappa$
is shown by solid curves.

\vspace{5mm}

Fig. 12.
Search for the upper boundary of the four-fermion excitation continuum.
Shown are the values of $\kappa_1$, $\kappa_2$, $\kappa_3$, $\kappa_4$
at which a maximum of $\omega/ \vert J\vert$ as given in (\ref{4.06}) occurs
at $\Omega=0.9\vert J\vert$
and $-\pi\leq\kappa\leq\pi$.
The dependences of $\kappa_1(=\kappa_2)$ 
and $\kappa_3(=\kappa_4)$ on $\kappa$ are shown by dashed curves,
the dependence of the maximal value of $\omega/ \vert J\vert$ on $\kappa$
is shown by solid curves.

\clearpage


\begin{thebibliography}{99}

\bibitem{01}
V. S. Viswanath and G. M\"{u}ller,
{\it The recursion method.
Application to many-body dynamics}
(Springer-Verlag, Berlin, Heidelberg, 1994)
(and references therein).

\bibitem{02}
M. Takahashi,
\newblock {\it Thermodynamics of one-dimensional solvable models}
\newblock (Cambridge University Press, Cambridge, 1999).

\bibitem{03}
V. E. Korepin, N. M. Bogoliubov, and A. G. Izergin,
\newblock {\it Quantum inverse scattering method and correlation functions}
\newblock (Cambridge University Press, Cambridge, 1993).

\bibitem{04}
D. C. Mattis,
\newblock {\it The many-body problem: An encyclopedia of exactly solved models in one dimension} 
\newblock (World Scientific, Singapore, 1993).

\bibitem{05}
G. M\"{u}ller and M. Karbach,
\newblock in {\it Frontiers of Neutron Scattering},
\newblock A. Furrer (Ed.)
\newblock (World Scientific, Singapore, 2000);
arXiv:cond-mat/0003076.

\bibitem{06}
J. Lorenzana and G. A. Sawatzky,
Phys. Rev. Lett. {\bf 74,} 1867 (1995);\\
J. Lorenzana and G. A. Sawatzky,
Phys. Rev. B {\bf 52,} 9576 (1995).

\bibitem{07}
H. Suzuura, H. Yasuhara, N. Nagaosa, and Y. Tokura,
Phys. Rev. Lett. {\bf 76,} 2579 (1996).

\bibitem{08}
Yongmin Yu, G. M\"{u}ller, and V. S. Viswanath,
Phys. Rev. B {\bf 54,} 9242 (1996).

\bibitem{09}
J. Lorenzana and R. Eder,
Phys. Rev. B {\bf 55,} R3358 (1997).

\bibitem{10}
R. Werner,
Phys. Rev. B {\bf 63,} 174416 (2001).

\bibitem{11}
E. Lieb, T. Schultz, and D. Mattis,
Ann. Phys. (N.Y.) {\bf 16,} 407 (1961).

\bibitem{12}
S. Katsura, 
Phys. Rev. {\bf 127,} 1508 (1962);
{\bf 129,} 2835 (1963).

\bibitem{13}
B. M. McCoy, E. Barouch, and D. B. Abraham,
Phys. Rev. A {\bf 4,} 2331 (1971).

\bibitem{14}
G. M\"uller and R. E. Shrock,
Phys. Rev. B {\bf 29,} 288 (1984).

\bibitem{15}
N. Kitanine, J. M. Maillet, and V. Terras,
Nucl. Phys. B {\bf 554,} 647 (1999).

\bibitem{16}
D. Biegel, M. Karbach, G. M\"{u}ller, and K. Wiele,
Phys. Rev. B {\bf 69,} 174404 (2004).

\bibitem{17}
J. Sato, M. Shiroishi, and M. Takahashi,
arXiv:cond-mat/0410102.

\bibitem{18}
G. M\"{u}ller, H. Thomas, H. Beck, and J. C. Bonner,
Phys. Rev. B {\bf 24,} 1429 (1981).

\bibitem{19}
J. H. Taylor and G. M\"{u}ller,
Physica A {\bf 130,} 1 (1985).

\bibitem{20}
T. Barnes,
Phys. Rev. B {\bf 67,} 024412 (2003).

\end{thebibliography}
\end{document}